\documentclass[aps,prb,superscriptaddress,twocolumn,nofootinbib,nobibnotes,floatfix,showpacs,reprint,longbibliography]{revtex4-1}
\usepackage{siunitx}
\usepackage[utf8]{inputenc}
\usepackage{amsmath,amssymb}
\usepackage{float} 
\usepackage{booktabs} 
\usepackage{bm} 
\usepackage{xcolor} 
\usepackage{xfrac} 
\usepackage{enumitem} 
\usepackage{soul} 
\usepackage[normalem]{ulem} 
\usepackage{subfigure}
\usepackage{multirow}
\usepackage{dcolumn} 
\usepackage{graphicx} 
\usepackage{hyperref} 
\hypersetup{
    unicode={true},
    colorlinks={true},
    linkcolor={blue},
    citecolor={blue},
    urlcolor={blue}
}

\begin{document}

\title{Topological Weyl Phase of an Ideal Spin-Gapless Semiconductor KCrSe } 

\author{Subhajit Mandal}
\affiliation{Department of Physics, Indian Institute of Technology Bombay, Mumbai 400076, India}

\author{Bishal Das}
\affiliation{Department of Physics, Indian Institute of Technology Bombay, Mumbai 400076, India}

\author{Himanshu Sharma}
\affiliation{Department of Physics, Indian Institute of Technology Bombay, Mumbai 400076, India}

\author{Satoru Hayami}
\affiliation{Graduate School of Science, Hokkaido University, Sapporo 060-0810, Japan}

\author{Aftab Alam}
\email{aftab@iitb.ac.in}
\affiliation{Department of Physics, Indian Institute of Technology Bombay, Mumbai 400076, India}

\begin{abstract}
The coexistence of topological and spin-polarized electronic states within a single material platform provides an attractive route toward emergent quantum phenomena and spintronic functionalities. However, materials simultaneously exhibiting spin-gapless semiconducting (SGS) behavior and Weyl semimetallicity remain exceedingly rare. Here, using first-principles calculations, we identify the half-Heusler compound KCrSe as an ideal spin-gapless Weyl semimetal. Transport calculations reveal a weak temperature dependence of the longitudinal conductivity and relatively small Seebeck coefficients, providing further evidence of its SGS nature.  KCrSe hosts a single pair of Weyl nodes---the minimum number permitted in a Weyl semimetal---located in close proximity to the Fermi level (E$_\text{F}$), resulting in exceptionally clean bulk and surface electronic spectra. The nontrivial Berry curvature associated with these Weyl nodes gives rise to sizable anomalous transport responses, including an anomalous Hall conductivity of $\sigma_{xy}^{A}\sim 90.76~\mathrm{S\,cm^{-1}}$ and an anomalous Nernst conductivity of $\alpha_{xy}^{A}\sim 0.15~\mathrm{A\,m^{-1}K^{-1}}$ at E$_\text{F}$, with substantially enhanced values at lower energies. The combination of an ideal Weyl topology, fully spin-polarized low-energy states, and finite anomalous transport establishes KCrSe as a promising platform for designing high-efficiency topological spintronic devices.
    
\end{abstract}

\date{\today}
\maketitle

{\it Introduction:} Spintronics has emerged as a promising alternative to conventional charge-based electronics by exploiting the spin degree of freedom of electrons for information processing and storage \cite{RevModPhys.76.323}. In this context, spin-gapless semiconductors (SGSs) constitute an unusual class of materials that combine semiconducting behavior with complete spin polarization. In an SGS, one spin channel exhibits a finite band gap, whereas the other possesses a gapless electronic spectrum at the Fermi level \cite{Yue2020SGS}. Such an electronic structure enables highly efficient spin transport with potentially negligible excitation energy, making SGSs attractive for next-generation spintronic devices.

Simultaneously, the discovery of topological quantum materials, including topological insulators and semimetals \cite{Moore2010Topological, RevModPhys.90.015001}, has fundamentally reshaped condensed matter physics. Among them, Weyl semimetals host pairs of topologically protected band-crossing points carrying opposite chirality, which act as sources and sinks of Berry curvature in momentum space \cite{Berry1984Phase}. These exotic quasiparticles give rise to remarkable transport phenomena such as the anomalous Hall effect, chiral anomaly, and unconventional magneto-transport responses. Weyl nodes emerge through the lifting of fourfold-degenerate Dirac crossings when either inversion symmetry ($\mathcal{P}$) or time-reversal symmetry ($\mathcal{T}$) is broken.

The coexistence of SGS behavior and nontrivial Weyl topology is particularly appealing because it combines fully spin-polarized transport with topologically protected electronic states. However, such materials remain exceedingly rare. Only a limited number of SGS compounds, including $\mathrm{CoFeMnSi}$ \cite{CoFeMnSi}, $\mathrm{CoFeCrGa}$ \cite{CoFeCrGa}, and Sr$_2$CuF$_6$ \cite{Sr2CuF6}, have been experimentally or theoretically identified to date. Even fewer systems simultaneously host SGS and Weyl characteristics, with $\mathrm{MnPO}_4$ \cite{MnPo4} and VTaNbAl \cite{VTaNbAl} being among the very few reported examples.

Heusler compounds provide a versatile platform for realizing such exotic quantum states owing to their tunable electronic structure, high Curie temperatures, and strong spin polarization \cite{GRAF20111,JAFARI2022110702}. In particular, half-Heusler alloys crystallizing in the noncentrosymmetric $F\bar{4}3m$ structure have recently attracted significant attention as candidates for topological phases \cite{Lin2010HalfHeusler,Liu2016LnPtBi,sgs_review}. Nevertheless, many experimentally studied Weyl semimetals, such as TaAs \cite{TaAs} and Co$_3$Sn$_2$S$_2$ \cite{Co3Sn2S2}, possess multiple pairs of Weyl nodes distributed far away from the Fermi level, leading to complicated bulk and surface spectra that often obscure their intrinsic topological signatures.

Here, using first-principles calculations, we identify half-Heusler KCrSe as an ideal spin-gapless Weyl semimetal. The compound exhibits a highly desirable combination of electronic and topological characteristics: (i) a single pair of Weyl nodes, representing the minimum number allowed in a Weyl semimetal, (ii) Weyl nodes located extremely close to the Fermi level, thereby maximizing their contribution to low-energy transport, and (iii) minimal trivial bulk states near the Fermi energy, resulting in exceptionally clean surface Fermi arcs. The nontrivial Berry curvature associated with the Weyl nodes generates sizable anomalous transport responses, including an anomalous Hall conductivity of $\sigma_{xy}^{A}\sim90.76~\mathrm{S\,cm^{-1}}$ and an anomalous Nernst conductivity of $\alpha_{xy}^{A}\sim0.15~\mathrm{A\,m^{-1}K^{-1}}$, comparable to or exceeding several previously reported SGS systems \cite{VTaNbAl,Mn2CoAl}. These findings establish KCrSe as a promising platform for investigating the interplay between spin-gapless electronic structure and topological Weyl physics.

{\it Computational Details:}
First principles calculations were performed within the framework of Kohn-Sham density functional theory (DFT) \cite{DFT1,DFT2} using the Vienna \textit{Ab-initio} Simulation Package (VASP) \cite{VASP1,VASP2,VASP3} based on the projector-augmented wave (PAW) \cite{PAW1,PAW2} method. The Perdew--Burke--Ernzerhof (PBE) \cite{PBE} functional within the generalized gradient approximation (GGA) was employed to describe the exchange-correlation interactions. Due to the presence of strong correlation between the $d$-electrons of Cr, self-interaction of Cr-$d$ electrons becomes significant and thus end up being delocalized.
To account for the localized nature of the Cr-3$d$ states, on-site Coulomb interactions were treated within the PBE+U formalism \cite{Dudarev1998}. The effective Hubbard parameter $U_{\text{eff}}$ = 4.8 eV was determined self-consistently using the linear-response approach \cite{PhysRevB.71.035105}.
The first Brillouin zone (BZ) was sampled using a $\Gamma$-centered $30 \times 30 \times 30$ $k$-mesh and kinetic energy cutoff of 340 eV, with the self-consistent electronic energy convergence criterion set to $10^{-6}$~eV. For band structure calculations, 100 $k$-points were chosen along each high-symmetry path. 

Maximally localized Wannier functions (MLWFs) \cite{MLWF1,MLWF2,MLWF3} were generated using \textsc{Wannier90} \cite{w90_1,w90_2,w90_3}, employing projections onto the $s$, $p$ orbitals of K, Se/Te atoms and $s$, $p$, $d$ orbitals of Cr atom to construct a tight-binding bulk Hamiltonian from pre-converged DFT results, where BZ was sampled using a $\Gamma$-centered $15\times15\times15$ $k$-mesh. We then performed nodal point search and chirality calculations on this bulk Wannier Hamiltonian.\cite{chirality_FHS}  The surface dispersion, Fermi surface, isoenergy contours were then obtained by applying the iterative Green's function method \cite{green1,green2,green3,green4} on a semi-infinite slab derived from the bulk Hamiltonian. We also calculated the anomalous Hall conductivity ($\sigma_{xy}$) and anomalous Nernst conductivity ($\alpha_{xy}$) using the following relations \cite{ahc,anc}:
\begin{gather}
    \sigma_{xy}^{A}(\varepsilon) = -\frac{e^2}{\hbar} \sum_{n} \int_{\mathrm{BZ}} \frac{d\mathbf{k}}{(2\pi)^3}\; f_{n}(\mathbf{k},\varepsilon) \;\Omega_{n,z}(\mathbf{k})
    \label{eq:AHC_3D_bands} \\
    \alpha_{xy}^{A}(\mu,T) = - \frac{1}{e} \int d\varepsilon \;\frac{\partial f_{n}(\mathbf{k},\varepsilon)}{\partial \mu} \;\sigma_{xy}^{A}(\varepsilon) \;\frac{\varepsilon - \mu}{T}
    \label{eq:ANC_mu_derivative}
\end{gather}
where, $\Omega_{n,z}$ is $z$-component of the Berry curvature for the $n$-th band, $f_n(\mathbf{k,\varepsilon}) = \left[1 + \exp{\left(\frac{E_{n}(\mathbf{k}) - \varepsilon}{k_{B}T}\right)}\right]^{-1}$ is the Fermi-Dirac distribution and $E_{n}(\mathbf{k})$ is the energy eigenvalue of $n$-th band, $\mathbf{k}$ being the crystal momentum. All the above post-processing of the tight-binding Wannier Hamiltonian was done via \textsc{WannierTools} \cite{Wtools}.


\begin{table}[t]
    \begin{ruledtabular}
        \caption{Relative total energies ($\Delta E_S$) per formula unit (f.u.) for different structural configurations of KCrSe. }
        \label{tab1}
        \begin{tabular}{c c c c c}
            \multirow{2}{*}{\shortstack{Structural \\ Configuration}} & \multicolumn{3}{c}{Wyckoff Position} & \multirow{2}{*}{$\Delta E_{S}$(meV/f.u.)} \\
            \cline{2-4}
            & 4a & 4b & 4c & \\
            \midrule
            Type I   & K & Cr & Se & 0 \\
            Type II  & K & Se & Cr & 209 \\
            Type III & Cr & Se & K & 1494 \\
        \end{tabular}
    \end{ruledtabular}
\end{table}


\begin{figure}[b]
    \centering
    \includegraphics[width=\linewidth]{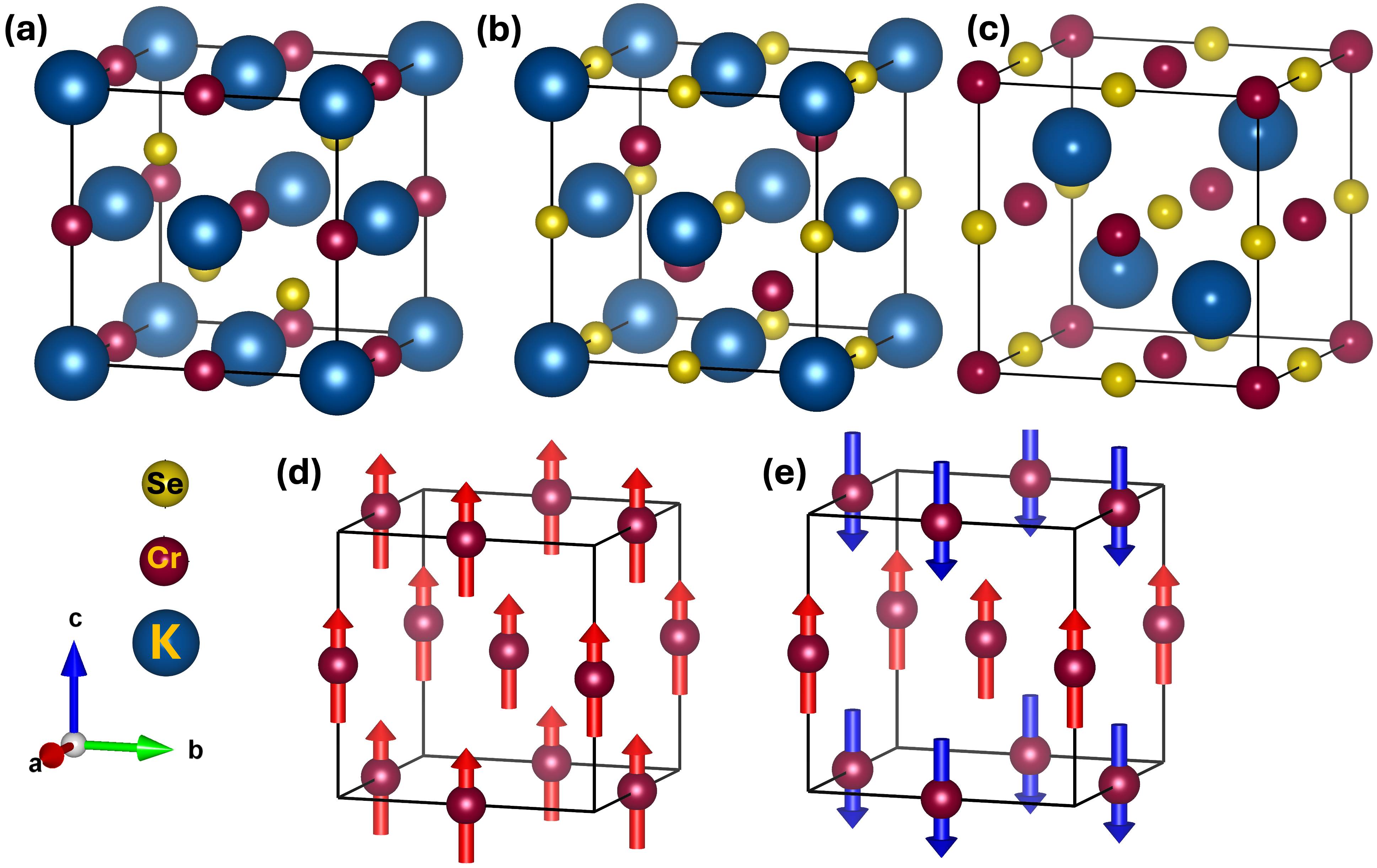}
    \caption{(a) Type I (b) Type II and (c) Type III configuration of KCrSe. (d) Ferromagnetic (FM) and (e) antiferromagnetic (AFM) configuration of KCrSe based on Type I structure. Only Cr atoms are displayed for clarity.}  
    \label{fig1}
\end{figure}

\begin{table}[t]
    \begin{ruledtabular}
        \label{tab:magnetic_configs}
        \setlength{\tabcolsep}{20pt} 
        \caption{ For KCrSe,  Relative total energies ($\Delta E_m$) per formula unit (f.u.)  for different magnetic configurations, referenced to the FM ground state of KCrSe.}
        \begin{tabular}{c c}
            Magnetic Configuration & $\Delta E_m$ (meV/f.u.) \\
            \hline
            \midrule
            FM           & 0   \\
            AFM          & 12  \\
            NM           & 31 \\
        \end{tabular}
    \end{ruledtabular}
\end{table}

{\it Crystal Structure and Magnetic Configurations:}
Heusler compounds have attracted sustained interest owing to their high Curie temperatures and remarkable versatility in hosting multiple correlated physical phenomena within a single material platform. Structurally, full-Heusler alloys crystallizing in the cubic $Fm\bar{3}m$ space group can be viewed as four interpenetrating face-centered-cubic sublattices occupying the Wyckoff positions {4a}$(0,0,0)$, {4c}$(0.25,0.25,0.25)$, {4b}$(0.5,0.5,0.5)$, and {4d}$(0.75,0.75,0.75)$. Among these, the A and C sites correspond to octahedral voids, whereas the B and D sites are tetrahedrally coordinated. Removing one of the tetrahedral sublattices lowers the symmetry to the noncentrosymmetric half-Heusler structure with space group $F\bar{4}3m$. For KCrSe, three nonequivalent atomic configurations are possible within this crystal structure, as illustrated in Fig.~\ref{fig1}(a-c), and their corresponding relative total energies are summarized in Table~\ref{tab1}. Type I turns out to be energetically the most favorable configuration. Further assessment of stability of KCrSe, together with those of other related compounds in this family, are reported elsewhere.\cite{KCrTe1,KCrSe_thermo}

To determine the magnetic ground state of KCrSe, we systematically investigated nonmagnetic (NM), ferromagnetic (FM), and antiferromagnetic (AFM) spin configurations based on Type I structure shown in Fig.~\ref{fig1}(d,e). The relative total energies presented in Table~\ref{tab:magnetic_configs} establish the FM phase as the energetically most favorable configuration. The calculated magnetic moment is approximately $4~\mu_B$ per Cr atom, originating predominantly from the Cr-$d$ states. Furthermore, the magnetocrystalline anisotropy energy (MAE) was determined by computing the ground-state energies corresponding to the (001), (110), and (111) magnetization axes. The energy difference between these magnetization axes lie within sub $\mu$eV, despite tightly converging the total energy calculations (30$^3$ k-mesh and 500 eV energy cut-off) in each case. Because these calculations yield effectively degenerate energies, we chose to align the magnetization along the (001) axis to simplify our subsequent topological studies. Varying the magnetization axis alters the orientation of the nodal points, while the underlying physics remains unchanged.



\begin{figure}[t]
    \centering
    \includegraphics[width=\linewidth]{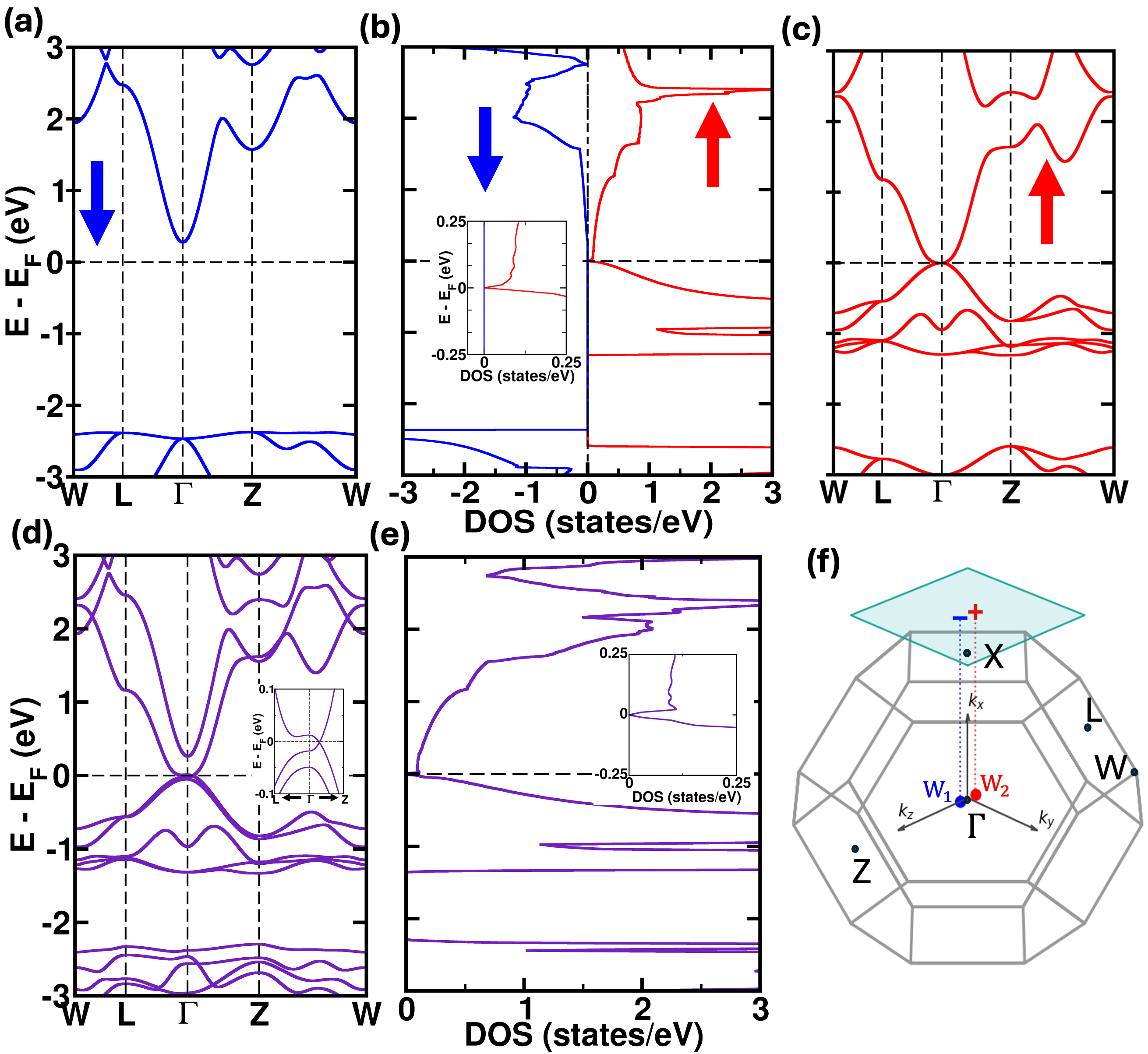}
    \caption{For KCrSe, (a,c) Band structure and (b) Density of States (DOS) for spin down ($\downarrow$) and up ($\uparrow$)  channels without SOC. (d) Band structure and (e) DOS with SOC. (f) Bulk Brillouin Zone marked with different high symmetry points and the (100) surface projected BZ.  }
    \label{fig2}
\end{figure}

{\it Bulk Electronic Structure:}
Figures~\ref{fig2}(a) and \ref{fig2}(c) display the spin-resolved electronic band structures for the minority- and majority-spin channels, respectively. A sizable band gap of approximately $2.5$~eV is observed in the minority-spin channel, whereas the majority-spin channel exhibits a vanishing band gap at the Fermi level, establishing the spin-gapless semiconducting (SGS) nature of KCrSe. Correspondingly, the majority-spin states dominate the low-energy electronic excitations, while the minority-spin channel remains insulating.
The SGS character is further confirmed by the spin-polarized density of states (DOS) shown in Fig.~\ref{fig2}(b). The DOS for the majority-spin channel vanishes precisely at the Fermi energy E$_F$ (see inset), consistent with a gapless semiconducting state, while the minority-spin channel exhibits a wide insulating gap. This coexistence of a gapless and an insulating spin channel represents the defining hallmark of an SGS.



\begin{table}[t]
    \begin{ruledtabular}
        \caption{Cartesian reciprocal coordinates and chiralities of Weyl nodes. }
        \begin{tabular}{c c c c c}
            Weyl Point & $k_x$~(\AA$^{-1}$) & $k_y$~(\AA$^{-1}$) & $k_z$~(\AA$^{-1}$) & Chirality \\
            \midrule
            W$_1$ & 0.0 & 0.0 & $+$0.068 & $-$1 \\
            W$_2$ & 0.0 & 0.0 & $-$0.068 & $+$1 \\
        \end{tabular}
        \label{Weyl_Nodes}
    \end{ruledtabular}
\end{table}

A closer inspection of the majority-spin band structure in Fig.~\ref{fig2}(c) reveals that two valence bands and one conduction band intersect at the $\Gamma$ point, forming a threefold-degenerate band crossing near the Fermi level. In the vicinity of this crossing, the electronic dispersion exhibits predominantly parabolic character. Upon inclusion of spin-orbit coupling (SOC) with magnetization oriented along the [001] direction, the overall electronic structure remains largely preserved, as shown in Fig.~\ref{fig2}(d). However, SOC lifts the threefold degeneracy at the $\Gamma$ point and generates a linearly dispersing band crossing along the $\Gamma$--Z direction (see inset of Fig.~\ref{fig2}(d)). Importantly, the relativistic effects do not alter the SGS character, as evidenced by the relativistic DOS shown in Fig.~\ref{fig2}(e), where the majority-spin DOS continues to vanish at E$_F$. As discussed in the following subsection, the SOC-induced linear crossing along $\Gamma$--Z corresponds to a Weyl node rather than an accidental band degeneracy.



{\it Weyl Nodes and Surface Dispersion:}
The ferromagnetic ground state of KCrSe breaks time-reversal symmetry ($\mathcal{T}$), thereby lifting Kramers degeneracy and restricting band crossings to at most twofold degeneracy. According to the Nielsen--Ninomiya theorem \cite{NIELSEN1981219}, such nodal crossings must occur in pairs with opposite chirality, giving rise to a Weyl semimetallic phase. As evident from Fig.~\ref{fig2}(d) and its inset, only two electronic bands intersect along the $\Gamma$--Z direction, resulting in a single Weyl node pair in the bulk Brillouin zone (the symmetry-related partner occurs along the -Z--$\Gamma$ direction). An extensive search throughout the entire Brillouin zone confirms the absence of any additional nodal crossings, establishing KCrSe as an ideal Weyl semimetal hosting the minimum possible number of Weyl node pairs.

\begin{figure}[b]
    \centering
    \includegraphics[width=\linewidth]{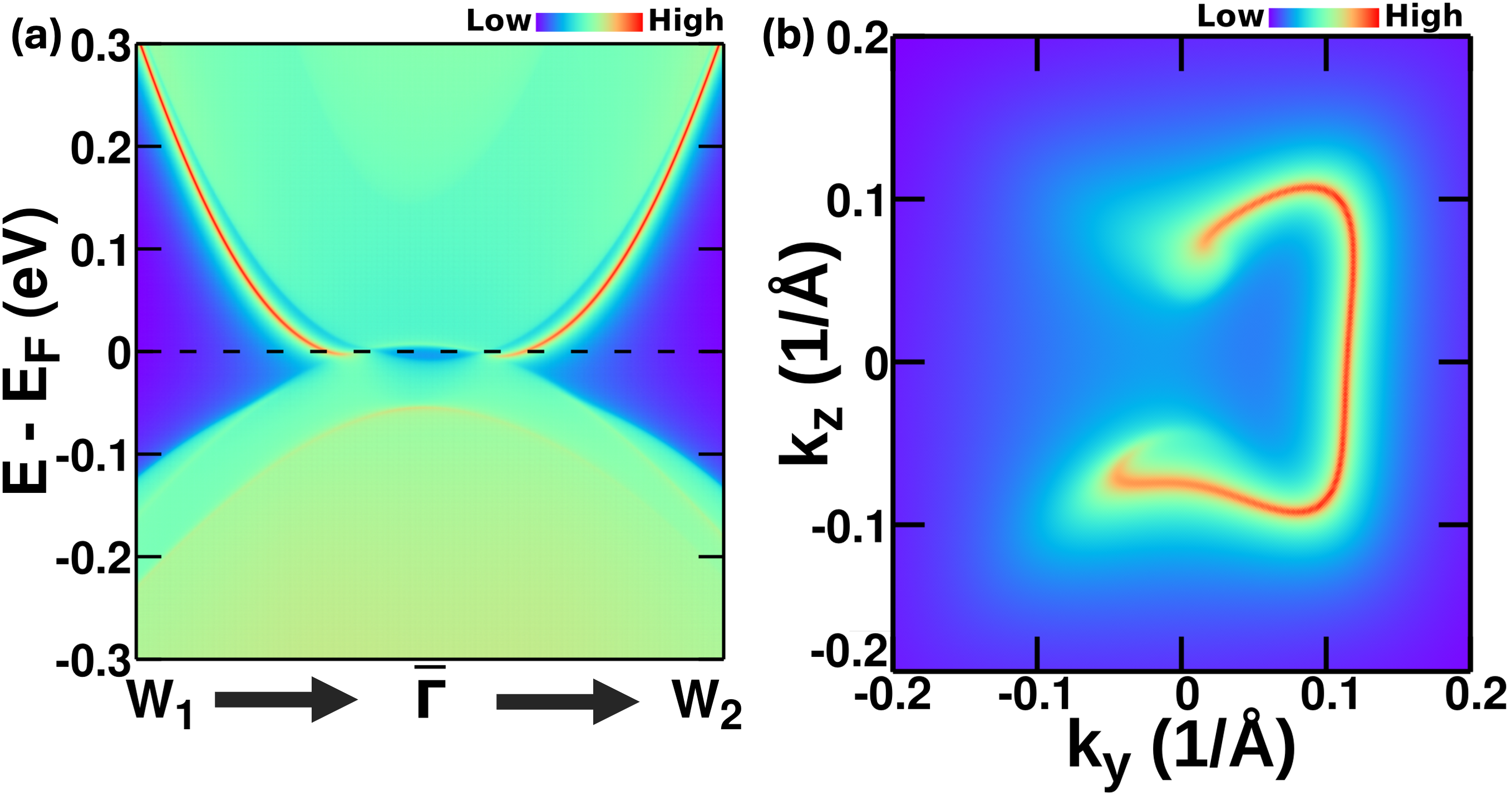}
    \caption{(a) Surface spectral dispersion projected onto the (100) surface of KCrSe. The dashed black line indicates the Fermi level (E$_F$). (b) Corresponding (100)-projected surface isoenergy contour (at an energy cut of 1 meV below E$_F$) showing the topological surface states and Fermi arc connectivity.}
    \label{fig3}
\end{figure}

The locations of the Weyl nodes within the bulk Brillouin zone and their projections onto the (100) surface Brillouin zone are illustrated in Fig.~\ref{fig2}(f) (by red and blue colored symbols), while their reciprocal-space coordinates and corresponding chiralities are listed in Table~\ref{Weyl_Nodes}. One of the hallmark signatures of Weyl semimetals is the emergence of topological surface states (TSSs) in the form of open Fermi arcs connecting Weyl nodes of opposite chirality. Owing to their spin-momentum locking and suppressed backscattering. These surface states are expected to exhibit highly efficient spin and charge transport characteristics.

\begin{figure}[t]
    \centering
    \includegraphics[width=\linewidth]{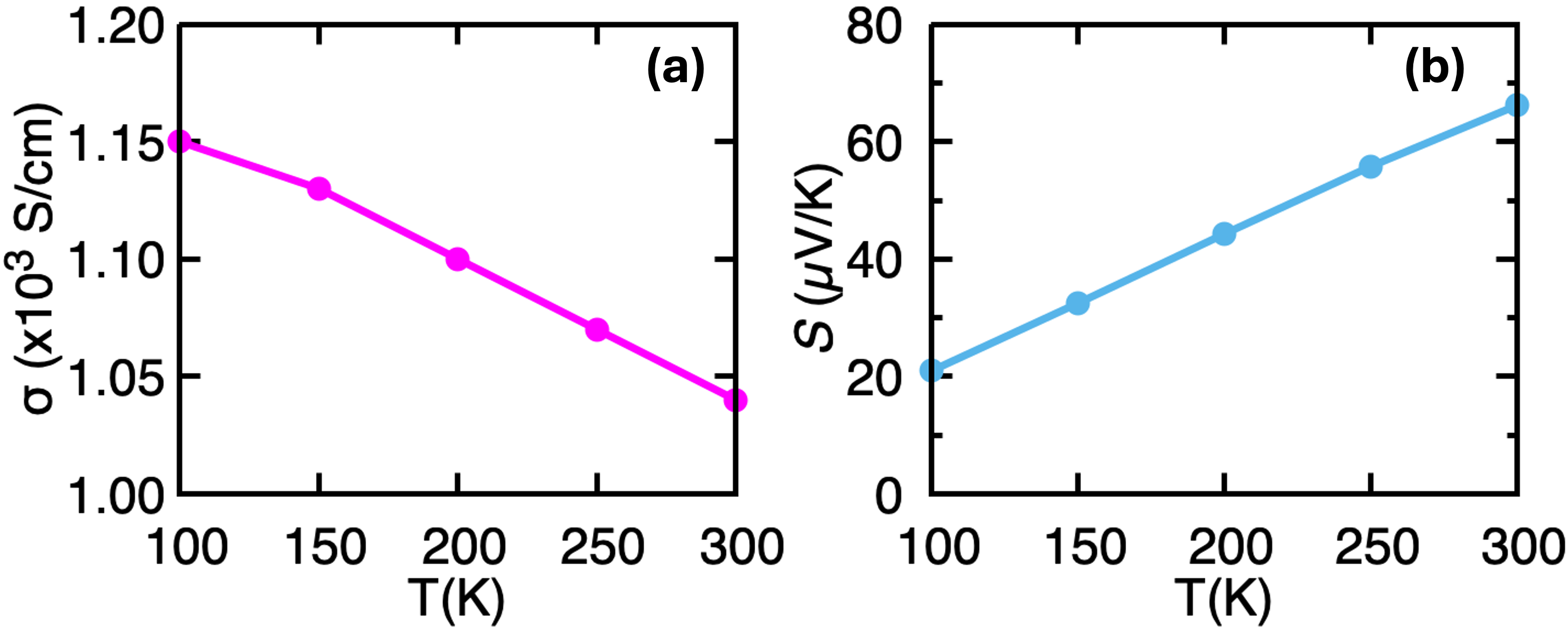}
    \caption{ Temperature (T) dependence of (a) electrical conductivity($\sigma$) and (b) Seebeck coefficient($S$) }
    \label{fig4}
\end{figure}

To probe the topological surface states in KCrSe, we constructed the cleaved (100) surface by breaking the $\mathrm{K}$--$\mathrm{Se}$ bonds. Since the bulk Weyl nodes are separated along the $k_z$ direction, their projections onto the (100) surface Brillouin zone remain distinct, enabling the appearance of surface Fermi arcs. Similar topological surface states are also expected on the (010) surface. Figures~\ref{fig3}(a) and \ref{fig3}(b) display the surface spectral function and the constant-energy contour evaluated at $15$~meV below the Fermi level, respectively, where the spectral weight obtained from the surface Green's function is represented by the color scale.

The low-intensity features in Fig.~\ref{fig3}(a) correspond to the projected bulk bands near the $\Gamma$ point and closely resemble the bulk dispersion shown in Fig.~\ref{fig2}(d). In contrast, the high-intensity states emerging from the projected Weyl nodes originate from topological surface states. These states connect the Weyl nodes of opposite chirality, forming a single open Fermi arc, as clearly visible in Fig.~\ref{fig3}(b).

{\it Thermoelectric Transport:}
We simulated the temperature ($T$) dependence of electrical conductivity ($\sigma$) and Seebeck coefficient ($S$) of KCrSe within a constant relaxation time $\tau$ = 10~fs, as shown in Fig.\ref{fig4}. The electrical conductivity remains nearly temperature independent and remains of the order of 10$^3$~S~cm$^{-1}$, consistent with the characteristics of other SGS materials \cite{CoFeCrGa,CoFeCrGa_x,CoFeMnSi,VTaNbAl,Mn2CoAl,Cr3Al}. The Seebeck coefficient remains positive across the entire temperature range, indicating hole dominated transport in this compound. As temperature increases, S rises to a maximum of $\sim$65~$\mu$V~K$^{-1}$ near 300 K. The relatively small magnitude of S is typical of spin gapless semiconducting systems\cite{Mn2CoAl,CoFeCrGa_x,Cr3Al} and reflects their vanishingly small effective energy gaps. Together with the electronic structure results, the weak temperature dependence of $\sigma$ and the relatively small Seebeck coefficient are consistent with the SGS character of KCrSe.

\begin{figure}[t]
    \centering
    \includegraphics[width=\linewidth]{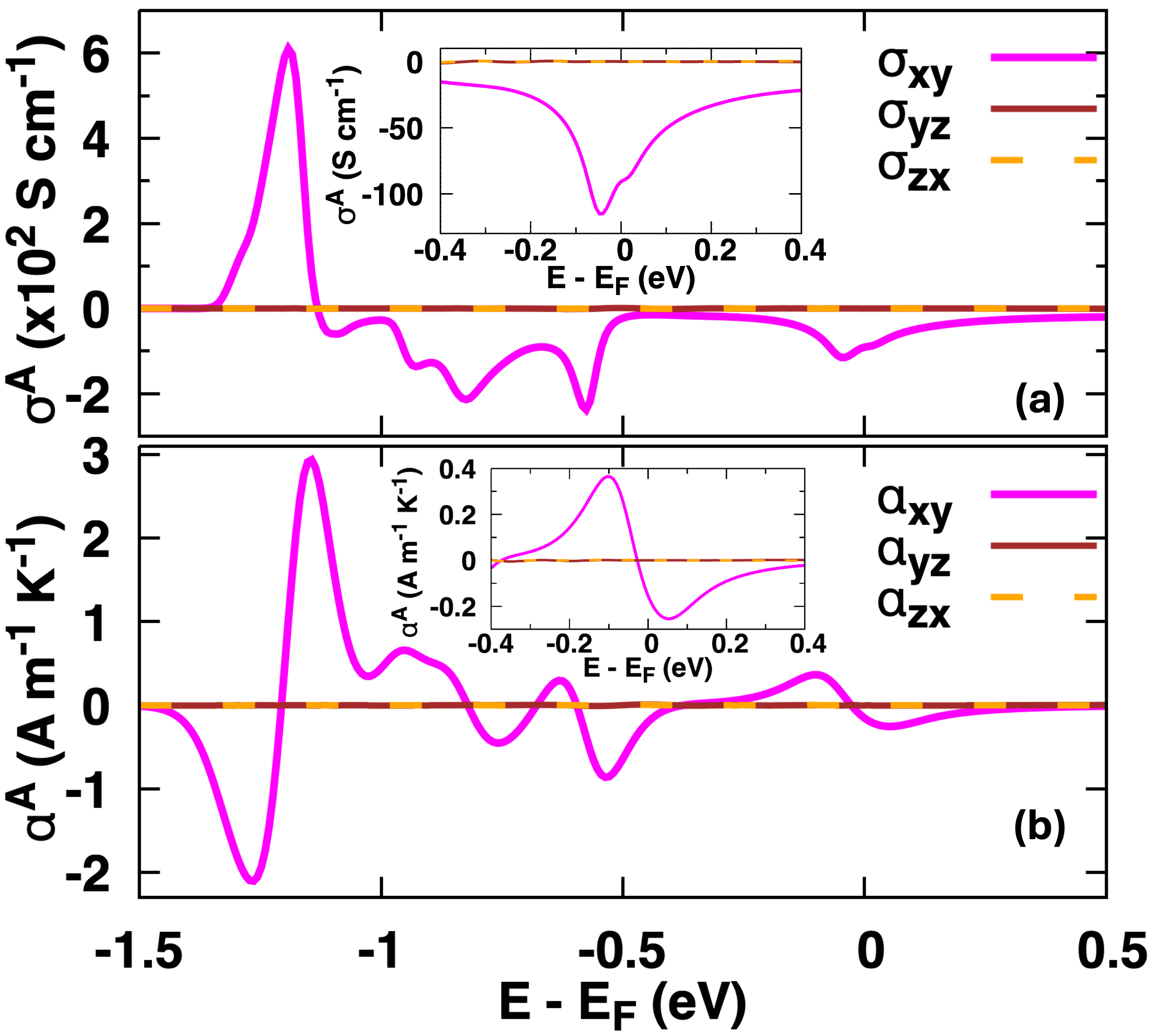}
    \caption{Variation of (a) anomalous Hall conductivity ($\sigma^{A}$) 
    and (b) anomalous Nernst conductivity ($\alpha^{A}$) of KCrSe with chemical potential $\mu$ = E - E$_F$ including SOC. Insets show zoomed-in plots around E$_F$. }
    \label{fig5}
\end{figure}


{\it Anomalous Transport:}
KCrSe realizes a ferromagnetic Weyl semimetallic phase hosting a single pair of Weyl nodes in close proximity to the Fermi level. The associated band crossings generate pronounced Berry curvature in momentum space, thereby giving rise to intrinsic anomalous transport responses. Figure~\ref{fig5}(a) presents the anomalous Hall conductivity (AHC) as a function of chemical potential, $\mu = E-E_F$, obtained by integrating the Berry curvature over the entire Brillouin zone using Eq.~\eqref{eq:AHC_3D_bands}. The corresponding anomalous Nernst conductivity (ANC) at $T=300$~K, evaluated using Eq.~\eqref{eq:ANC_mu_derivative}, is shown in Fig.~\ref{fig5}(b).

The magnetic moments are aligned along the [001] direction, which breaks the mirror symmetry of the $xy$ plane and lifts the band degeneracies within the $k_x$--$k_y$ plane. Consequently, the Weyl nodes emerge along the $\Gamma$--Z direction, resulting in a dominant $z$ component of the Berry curvature. As a result, only the transverse components $\sigma^{A}_{xy}$ and $\alpha^{A}_{xy}$ exhibit appreciable magnitudes, while all other tensor components remain negligible throughout the investigated energy window.
Both $\sigma^{A}_{xy}$ and $\alpha^{A}_{xy}$ remain relatively small over a broad energy range extending from approximately $-0.5$ to $0.5$~eV, except in the vicinity of the Fermi level where the Weyl nodes are located. At E$_F$, the Weyl-induced Berry curvature yields an anomalous Hall conductivity of approximately $90.76~\mathrm{S\,cm^{-1}}$ and an anomalous Nernst conductivity of about $0.15~\mathrm{A\,m^{-1}K^{-1}}$ at $300$~K (see insets of Figs.~\ref{fig5}(a) and ~\ref{fig5}(b) respectively). 

Additionally, in weak spin-orbit-coupled systems, the magnitude of the anomalous Hall response is strongly influenced by the degree of spin polarization. The total anomalous Hall conductivity can be expressed as the sum of the independent spin-channel contributions,
\[
\sigma_{xy}^{A,\mathrm{total}}=\sigma_{xy}^{A,\uparrow}+\sigma_{xy}^{A,\downarrow}.
\]
As evident from the non-relativistic band structure shown in Figs.~\ref{fig1}(a) and \ref{fig1}(c), the electronic states near the Fermi level are fully spin polarized and originate exclusively from the majority-spin channel. Consequently, the anomalous Hall response around E$_F$ is primarily governed by the spin-up bands, while the minority-spin contribution remains negligible. This limits the overall magnitude of the AHC in the vicinity of the Fermi level despite the presence of Weyl-mediated Berry curvature.


{\it Conclusions:}
In summary, using first-principles calculations, we identify KCrSe as an ideal spin-gapless semiconductor simultaneously hosting a topologically nontrivial Weyl semimetallic phase. The transport calculations, showing weak temperature-dependent longitudinal conductivity and relatively low magnitude of Seebeck coefficients, further support its SGS behavior. The electronic structure exhibits only a single pair of Weyl nodes located near the Fermi level, realizing an ideal Weyl topology with minimal bulk-state interference. The coexistence of fully spin-polarized gapless excitations and Weyl fermions makes KCrSe an attractive platform for exploring topological spintronic phenomena.
The intrinsic SGS character enables tunability of the low-energy electronic states through external perturbations such as chemical doping or strain. Moreover, the minimal number of Weyl nodes gives rise to exceptionally clean surface spectra with well-defined topological Fermi arcs favorable for low-dissipation transport. The Weyl-mediated Berry curvature produces sizable anomalous transport responses, including an anomalous Hall conductivity of approximately -$90.76~\mathrm{S\,cm^{-1}}$ and an anomalous Nernst conductivity of about -$0.15~\mathrm{A\,m^{-1}K^{-1}}$ at the Fermi level. Our findings establish KCrSe as a promising platform for investigating intrinsic Weyl physics and designing next-generation topological spintronic devices.

{\it Acknowledgements:}
SM acknowledges the High Performance Computing Cluster facility 
\textquoteleft RUDRA\textquoteright{} at IIT Bombay to carry out few calculations reported in this work.

\bibliographystyle{unsrt}
\bibliography{ref}

\end{document}